# CMOS-compatible titanium nitride for on-chip plasmonic Schottky photodetector

Jacek Gosciniak[1,*], Fatih B. Atar[2], Brian Corbett[2], and Mahmoud Rasras[1]

[1]New York University Abu Dhabi, Saadiyat Island, PO Box 129188, Abu Dhabi, UAE
[2]Tyndall National Institute, Lee Maltings, Cork, Ireland
[*]jg5648@nyu.edu

**ABSTRACT**
Here, we propose titanium nitride (TiN) as an alternative plasmonic material for an on-chip silicon plasmonic Schottky photodetector that is based on an internal photoemission process and operating at telecom wavelengths. The examined structure employs an asymmetric metal-semiconductor-metal waveguide structure with one of the electrodes being gold and the second either gold, titanium or titanium nitride. Apart from the excellent optical properties desired for this type of photodetector such as high absorption losses and reasonably high real part of the permittivity, titanium nitride is a CMOS-compatible material that enables easy integration with existing CMOS technology. For the first time, we find a Schottky barrier height of 0.67 eV for titanium nitride on p-doped silicon, which is very close to the optimal value of 0.69 eV. This value ensures very high signal-to-noise ratio of the photodetector operating at a wavelength of 1550 nm. Additionally, TiN provides shorter penetration depth of the mode into metal compared to Ti, which enhances transmission probability of hot electrons to a semiconductor and gives rise to responsivity enhancement.

## I. INTRODUCTION

Photodetectors (PDs) are one of the basic building blocks of an optoelectronic link that converts light into an electrical signal. On-chip monolithic optoelectronic integration requires the development of CMOS-compatible PDs that operate in the telecom wavelengths (1.1 - 1.7 μm) [1-3]. Although sensitivity is the most important attribute for photodetectors in long distance communications, for short distance interconnects the most critical factor is the total energy dissipated per bit. The optical energy received at a photodetector is directly related to the transmitter optical output power and the total link loss power budget, which includes total link attenuation, coupling losses and eventually, a power margin. Hence, for 10 fJ/bit transmitted optical energies, the received optical energy would be 1 fJ/bit [3]. Thus, minimizing the optical losses at the photodetector is crucial for the overall performance of the system.

Photodetectors usually operate on the basis of the photoelectric effect or exhibit an electrical resistance dependent on the incident radiation. The operational principal is based on the absorption of photons and the subsequent separation of photogenerated charge carriers – electron-hole (e - h) pairs [4]. However, they suffer from low efficiency either because the near-infrared (NIR) photons energies (0.79 - 0.95 eV) are not sufficient to overcome the Si bandgap (1.12 eV) or due to small PD detection area in the case of Ge-based photodetectors (bandgap 0.67 eV) [4]. An alternative approach utilizes the intrinsic absorption of metal for photodetection that is accomplished by internal photoemission (IPE) in a Schottky diode [5-10]. In this configuration, the photoexcited ("hot") carriers from the metal are emitted to the semiconductor/insulator over a potential $\Phi_B$, called the Schottky barrier, that exists at the metal−semiconductor interface.

In the semiconductor/insulator, the injected carriers are accelerated by an electric field present in the depletion region of a Schottky diode and then collected as a photocurrent at the external electrical contacts. Usually, a Schottky barrier is lower than the bandgaps of most semiconductors, thus allowing a photodetection of near-infrared (NIR) photons with energy hν > $\varphi_B$. The process of photon-induced emission of electrons from metals and its collection was described by Spacer [11, 12]. It is based on the Fowler proposal and consists of a three-step model: (1) generation of hot electrons in the metal through the absorption of photons, (2) diffusion of a portion of the hot electrons to the metal-semiconductor/insulator interface before thermalization, and (3) injection of hot electrons with sufficient energy and correct momentum into the conduction band of the semiconductor/insulator through an internal photoemission.

To enhance the efficiency of the IPE process it is desirable to confine the optical power at the boundary between materials forming the Schottky barrier. This allows for increasing the interaction of light with the metal in a very close vicinity of the interface where the photoemission process takes place. The solution for this is well known and it is called a surface plasmon polariton (SPP). The SPPs are guiding optical surface waves propagating along the boundary between the metal and dielectric with a maximum field located at this interface and decaying exponentially in both media [13, 14]. One of the main advantages of SPP emanates from the fact that it is not diffraction limited and it enables a tight confinement of the optical field to strongly subwavelength dimensions. SPP offers a long interaction length between the propagating mode and the photodetector, thus it allows a larger portion of the optical energy to be absorbed nearby the Schottky barrier.

Plasmonic photodetectors are suggested by ITRS as an alternative technology to overcome the foreseeable scaling limitations of conventional components [15]. It offers multiple advantages such as high integration densities, low device capacitance allowing for higher bandwidth operation, and ultra-low energies to operate. SPP can decay either radiatively via emission of photons or non-radiatively through the generation of excited carriers, so-called hot carriers. These photo-excited hot carriers are able to overcome the potential barrier between the metal and semiconductor/insulator, which leads to a light-induced charge separation and hence a measurable current. Furthermore, the potential barrier can be overcome either directly or through quantum mechanical tunnel effects with the probability depending on the barrier width and height as well as the charge carrier energy.

## II. PHOTODETECTOR DESIGN

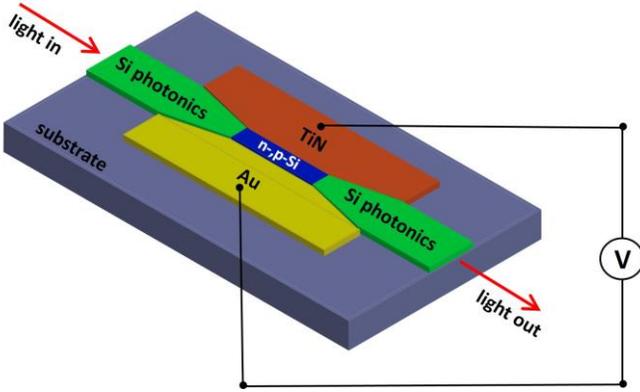

Fig. 1. Schematic of the asymmetric metal-semiconductor-metal (MSM) waveguide structure implemented for photodetection with a light coupled from the Si waveguide to the MSM junction, which is biased with the external voltage.

The considered arrangement consists of a metal-semiconductor-metal (MSM) structure with the semiconductor Si placed between the metals (see Fig. 1), similar to the one presented in reference [5]. For this plasmonic waveguide structure, the electric field of the propagating mode reaches its maximum at both metal-semiconductor interfaces and decays exponentially into the semiconductor, with the electromagnetic energy located partially in the metal and semiconductor. The amount of energy in the metal and dielectric depends on the material's optical properties and the waveguide geometry. Furthermore, the penetration depth of the electromagnetic field into the semiconductor depends significantly on the permittivity of the semiconductor and the plasmonic waveguide configuration. In contrast, the penetration depth of light into a metal, the skin depth, depends on the metal's optical properties [13, 14]. Thus, a large negative real permittivity, which is a consequence of larger plasma frequency or larger carrier concentration, gives a small penetration into the metal, while a small imaginary permittivity leads to lower absorption in the metal [13, 14]. Based on this it can be deduced that field penetration into the metal influences the trade-off between confinement and propagation losses – i.e., the less light inside the metal and more inside the dielectric, the smaller the absorption and confinement.

Here we examine TiN as a replacement for Ti as it shows much better plasmonic behavior and, in consequence, it is able to confine an absorbed energy much closer to metal-semiconductor interface. As a result, the hot carriers have much higher probability to be transfer through a Schottky barrier. Simultaneously, good plasmonic properties comparable with this of Au makes TiN an ideal candidate as a replacement for Au and development of CMOS-compatible photodetectors.

## III. OPERATION PRINCIPLE

The operation principle is based on the intrinsic absorption of metals that is accomplished by internal photoemission (IPE) [5-10]. The absorbed photons in the metal creates hot carriers that are transmitted across a potential barrier at the metal-semiconductor (MS) interfaces. Plasmonics are ideally suited for realizing such photodetectors as a metal stripe can be used both to support the plasmonic mode and as electrodes, with the mode field strongly localized at the MS interface where it reaches its maximum. Thus, light is perfectly concentrated in the region where absorption leads to the highest generation rate of photo-electrons. To minimize a dark current in the presented MSM configuration, one of the metal electrodes needs to be more absorbant compared to the other. Hence, an asymmetric MSM structure is highly desired where the light is mostly absorbed in one MS interface.

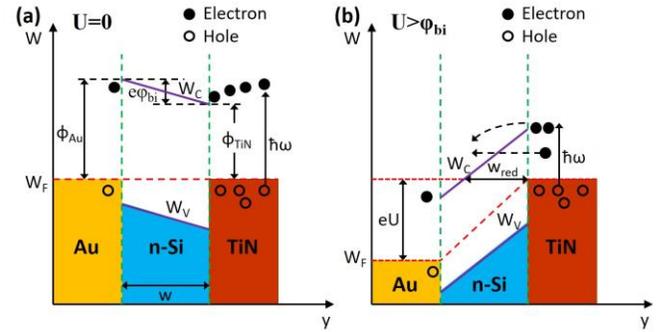

Fig. 2. Energy band diagram of the Au-Si-TiN junction in (a) thermal equilibrium, no bias voltage and (b) nonequilibrium under applied forward bias voltage V, positive in the Au → TiN direction.

The band diagram of an asymmetric MSM, Au-Si-TiN, photodetector junction is sketched in Fig. 2 with the plasmonic mode guided between the metal contacts (Fig. 1). In this junction, a SPP mode mostly dissipates its energy at the Si-TiN interface, and the absorbed photons create hot electrons in the TiN metal [24, 25]. When the maximum carrier energies exceed the Fermi energy, the hot electrons have an increased probability of crossing the potential barrier at the TiN-Si interface. Also, if the Schottky barrier height at the second MS interface (Au-Si) exceeds that of TiN-Si interface, the built-in potential difference $\varphi_{bi}$ across the silicon core impedes electron photoemission. Therefore, no significant current flow can be observed. For an applied voltage to the Au-TiN electrodes, when a positive potential at the Au electrode exceeds $\varphi_{bi}$, the photoemission from TiN is enabled.

### A. Optical properties

In this paper we examine the impact of different metal materials such as gold (Au), titanium (Ti) and titanium nitride (TiN) on the performances of the photodetector. Gold has been the metal of choice for most plasmonic components [13, 14]. However, it exhibits low melting temperature, low mechanical durability, high surface energies, and incompatibility with standard complementary metal-oxide-semiconductor (CMOS) fabrication [16, 17]. These major drawbacks have inhibited the full realization of many applications. In contrast, Ti and TiN are CMOS compatible, and are characterized by high melting temperature, extreme mechanical durability and low surface energy [16]. Furthermore, TiN shows optical properties similar to those of gold which makes it attractive for many applications [16-19]. In our MSM photodetector arrangement, we keep the left electrode to be Au, while the second electrode changes between Au (symmetric MSM) [25], and Ti and TiN (asymmetric MSM) [26].

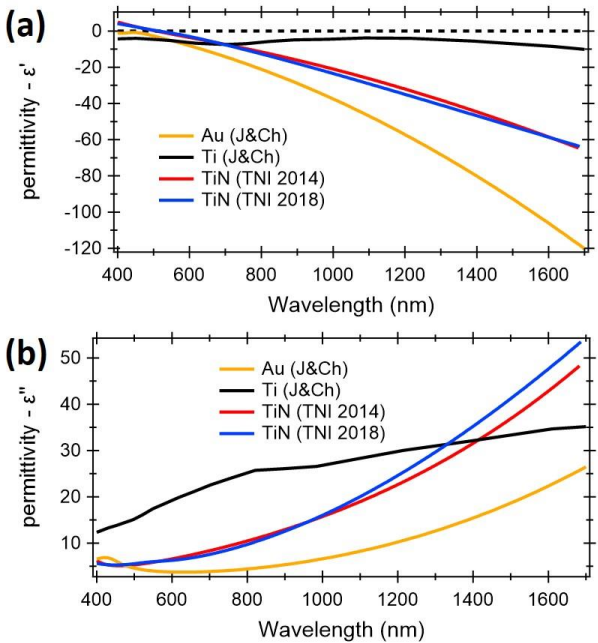

Fig. 3. Dielectric function of Au, Ti and TiN – (a) real and (b) imaginary part. For TiN the optical properties were measured for 50 nm thick film deposited on SiO2. Dielectric functions of Au and Ti were taken from a literature [18, 19].

The TiN was fabricated under different conditions and as a result it shows slightly different optical properties (Fig. 3). As observed, Au shows the highest negative real part of permittivity, while the imaginary part is the lowest among the presented materials. In comparison, Ti shows a very small negative real part of permittivity while the imaginary part is higher compared to Au. In the case of TiN, it shows a similar real part as Au, while the imaginary part of permittivity for the TiN sample TNI 2018 is higher for longer wavelengths. It is a consequence of slightly higher oxygen incorporation during a fabrication process.

*B. Electrical properties*

To characterize the electrical properties of TiN-Si contacts, we measured a current-voltage (I-V) characteristic of the Schottky contact for both n-doped (n=2-4 Ω·cm) and p-doped (n=10-20 Ω·cm) silicon. The TiN thickness was kept constant at h=50 nm for which a sheet resistance was measured at 50 Ω/sq. (ρ=2.5·104 Ω·cm) while its diameter changed from d=100 μm to d=200 μm.

A ring-shaped Au structure was formed the top contact on the TiN/p-Si device. The substrate was used as the bottom contact (inset in Fig. 4b). Current-voltage measurements were done by using a Keithley Sourcemeter.

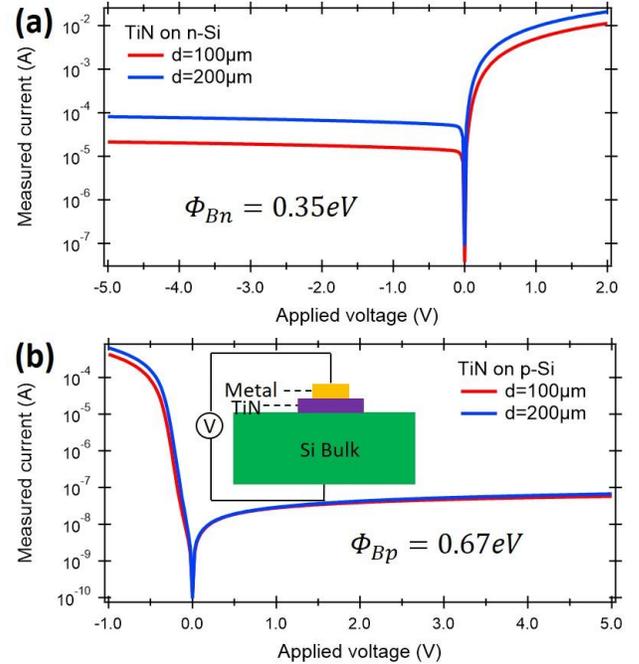

Fig. 4. I-V characteristics of the fabricated TiN-Si junction (inset in Fig. 4(b)) for different TiN contact areas – diameter d=100 μm and d=200 μm. The Schottky barrier heights of $\Phi_B$=0.35 eV and $\Phi_B$=0.67 eV for (a) n-doped and (b) p-doped Si respectively.

For a Schottky photodetector, the barrier height is a very important parameter. Values of the barrier heights for Au and Ti were taken from the literature [5, 7, 20] while for TiN it was based on our I-V measurements (Fig. 4) [9]. The dark current of Schottky diode is expressed by

$$I = SA^*T^2 exp\left(\frac{e\Phi_B}{kT}\right)\left[exp\left(\frac{eV}{kT}\right) - 1\right]$$

where S is the contact area, A* is the effective Richardson constant, $\Phi_B$ Schottky barrier height, and V is the applied voltage.

As it can be observed from Fig. 4, the smaller contact area results in a smaller dark current. However, this behavior is better pronounced for the TiN on n-doped Si (Fig. 4a) as a result of lower Schottky barrier height compared TiN on p-doped Si so more carriers can flow from TiN to Si. For p-Si (Fig. 4b) the device shows expected rectifying behavior with the forward bias region limited by the series resistance of the contact ($R_S$=822 Ω) and dark current in order of 8.1 nA for reverse bias of 0.1 V. The ideality factor for this device was calculated at n=1.3. On the contrary, for n-Si (Fig. 4a) the series resistance of the contact was calculated at RS=110 Ω and dark current of around 0.46 μA for a bias voltage of −0.1 V. The ideality factor was calculated again at 1.3. The Schottky barriers deducted from the curves were $\Phi_{TiN}$=0.35 eV for n-Si and $\Phi_{TiN}$=0.67 eV for p-Si. The insert in Fig. 4(b) shows the measurement setup. The barrier height for Au on n-Si and p-Si -type substrates were taken at $\Phi_{Au}$=0.8 eV and $\Phi_{Au}$=0.32 eV respectively, while for Ti were taken at $\Phi_{Ti}$=0.5 eV for n-Si and $\Phi_{Ti}$=0.61 eV for p-Si.

Thus, for n-Si core based MSM structure, the calculated, established built-in potential difference across the core is $\varphi_{bi}$=0.3 eV and $\varphi_{bi}$=0.45 eV for Au-Si-Ti and Au-Si-TiN, respectively. For p-doped Si, the corresponding potential difference is $\varphi_{bi}$=-0.29 eV for Au-Si-Ti and $\varphi_{bi}$=-0.35 eV for Au-Si-TiN. A negative potential sign means a lower Schottky barrier at the Au interface compared to the second interface – Ti-Si or TiN-Si. It should be mentioned here that the Schottky barrier height between Au and p-doped Si differs greatly from the data presented in reference [5] where the Schottky barrier height was taken as $\Phi_{Au}$=0.82 eV. However, even references provided in this paper suggest this value refers to n-doped Si rather than p-Si [21]. Consequently, the built-in potential difference of $\varphi_{bi}$=0.2 eV in the Au-(p-Si)-Ti used in paper [5] is not proper. However, the presented results constitute an excellent step in the realization of future waveguide-integrated photodetectors.

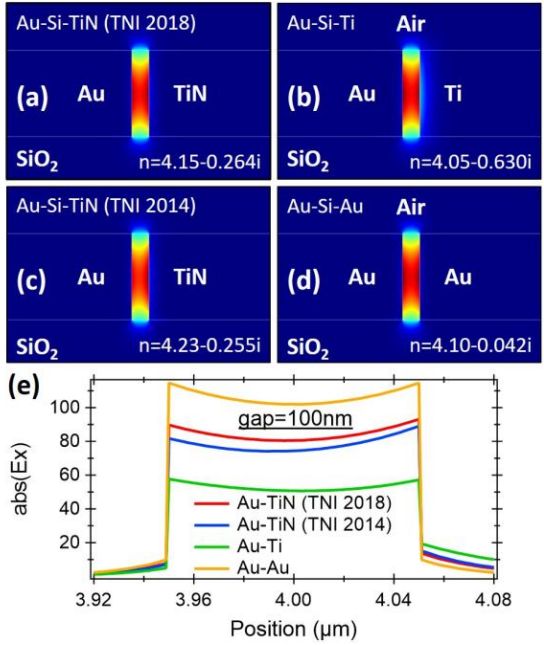

Fig. 5. Longitudinal component of the electric field inside an Si gap and into the metals for a gap width of w=100 nm and for different metal-semiconductor-metal materials: (a) Au-Si-TiN (TNI 2018), (b) Au-Si-Ti, (c) Au-Si-TiN (TNI 2014) and (d) Au-Si-Au, respectively. The corresponding electric field magnitude as a function of position into the gap and metals, for all the metal combinations, is shown in (e).

## IV. MSM ARRANGEMENT

When light is coupled to the MSM plasmonic waveguide it dissipates its energy at both metal-semiconductor interfaces. The amount of power absorbed at a given interface depends on the metal's optical properties – the larger the magnitude of the imaginary part of the complex permittivity, the larger the absorption. In our calculations, the complex permittivities of Au, Ti and TiN [18, 19] were taken at the wavelength of 1550 nm (Fig. 3). The MSM waveguide arrangement, where the Si core is placed between the metals, is similar to the structure presented in reference [5]. Figure 5 and 6 illustrates the longitudinal component of the electric field inside a Si gap and into the metals for a gap width of w=100 nm (Fig. 5) and w=200 nm (Fig. 6) for different metal-semiconductor-metal materials: Au-Si-TiN (TNI 2018), Au-Si-Ti, Au-Si-TiN (TNI 2014) and Au-Si-Au, respectively. The corresponding electric field profile as a function of position, for all the metal combinations, is shown in (i) and (j). This allows for observation of the influence of both material properties and gap width on the electric field's distribution inside the gap, as well as the absorption at the metal-semiconductor interfaces. Figures 7 and 8 show results for the same metal combination but for a wider gap of 500 nm. As mentioned in the "Photodetector design" section, the amount of the energy in the metal and dielectric depends on the material's optical properties and the waveguide geometry. A large negative real permittivity of a metal gives small penetration into the metal, while a small imaginary permittivity leads to lower losses, hence absorption. The penetration depth into the metal is defined mostly by the longitudinal component of the electric field. For waveguide-integrated photodetectors, the smaller penetration depth into the metal means more hot carriers can participate in the transition to the semiconductor, enhancing the photocurrent. As the penetration depth in the metal increases, the transition probability decreases. Hot electrons generated far from the metal-semiconductor interface can lose their energy through scattering which reduces the transition probability to the semiconductor. Thus, for photodetection purposes the main objective is to achieve high absorption as close as possible to the metal-semiconductor interface.

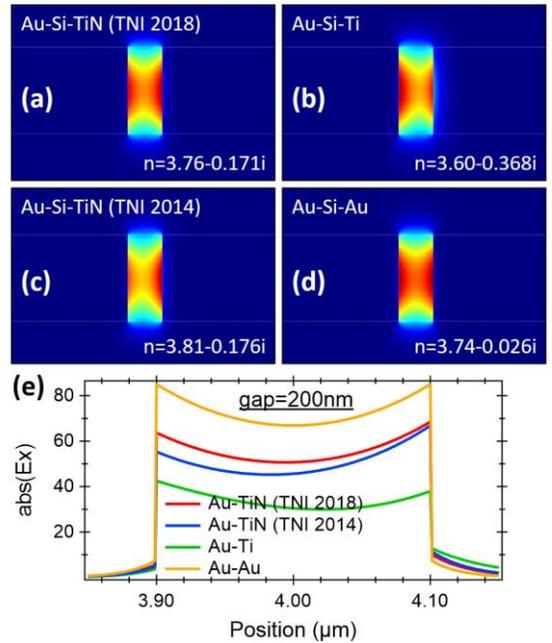

Fig. 6. Longitudinal component of the electric field inside an Si gap and into the metals for a gap width of w=200 nm for different metal-semiconductor-metal materials: (a) Au-Si-TiN (TNI 2018), (b) Au-Si-Ti, (c) Au-Si-TiN (TNI 2014) and (d) Au-Si-Au, respectively. The corresponding electric field magnitude as a function of position into the gap and metals, for all the metal combinations, is shown in (e).

For the Au-Si-Au structure, the electric field distribution both in the gap and in the metal is symmetric, which limits its application in a photodetector. Furthermore, absorption in the metal is very small, thus not a significant amount of hot carriers are generated. In comparison, the absorption in the metal is highly enhanced when Ti-Si or TiN-Si interfaces are used in Au-Si-Ti or Au-Si-TiN structures, respectively.

The electric field magnitude into a metal-semiconductor interface for a gap width of g=100 nm and for Ti-Si interface reaches 19.3 (a.u.) while for the TiN-Si interface it exceeds 15.1 (a.u.) and 13.6 (a.u.). Simultaneously, the absorption depth for Ti is much longer and even after 50 nm the electric field into a metal reaches a magnitude of 6.3 (a.u.) while for TiN this value is achieved after only 25 nm.

As mentioned earlier, hot carriers generated far from the metal-semiconductor interface have a low probability of participating in a transition to a semiconductor. Taking into account an electron mean free path into Ti and TiN that is in the range of 50 nm [27], we expect a low probability of transition for the hot carriers generated above this distance. It is worth noticing that the imaginary part of the mode effective index for the Au-Si-Ti structure ($n_{im}$=0.630·i) is more than 2.5 times higher than the imaginary part of the Au-Si-TiN structure ($n_{im}$=0.264·i and $n_{im}$=0.255·i) (Fig. 5). Thus, making the Au-Si-

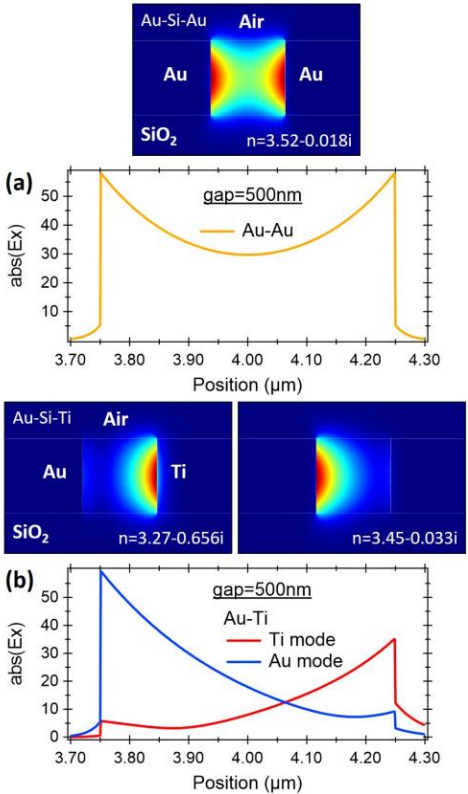

Fig. 7. Longitudinal component of the electric field inside a Si gap and into the metals for a gap width of w=500 nm and for (a) symmetric Au-Si-Au structure, and (b) asymmetric Au-Si-Ti structure.

TiN photodetector twice as long we can achieve even better absorption into the metal and, simultaneously, more hot carriers will be generated close to the TiN-Si interface that can participate in a photocurrent generation. As reported in references [29], the bandwidth of the MSM photodetectors is usually not limited by the RC time but rather by the carrier transit time. Thus, making a photodetector even 2.5 times longer will not influence its bandwidth.

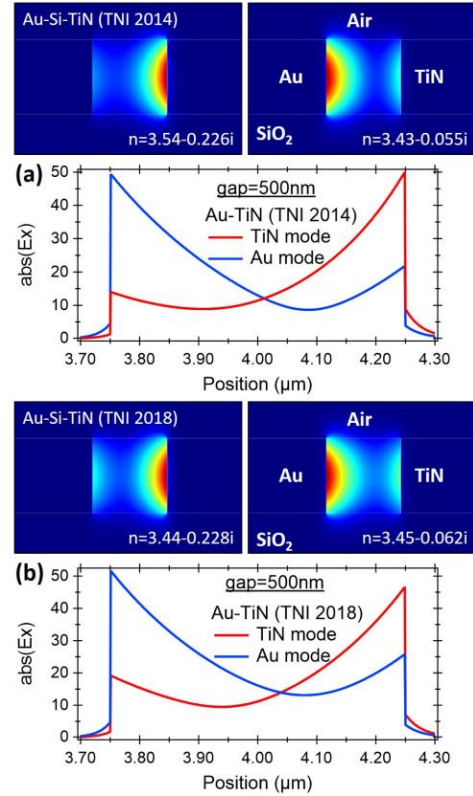

Fig. 8. Longitudinal component of the electric field inside a Si gap and into the metals for a gap width of w=500 nm and for asymmetric Au-Si-TiN structures with TiN obtained under slightly different deposition conditions – (a) sample TiN 2014 and (b) sample TiN 2018. Corresponding permittivities from Fig. 3.

Even more pronounced behavior is observed for the wider gap of g=200 nm (Fig. 6). The magnitude of the electric field at the Ti-Si interface for the Au-Si-Ti structure reaches 12.8 (a.u.) and drops to 4.3 (a.u.) after 50 nm inside the Ti. In comparison, the magnitude of the electric field at the TiN-Si interface reaches 11.4 (a.u.) and drops to 2 (a.u.) after the same distance of 50 nm inside the TiN. Furthermore, the imaginary part of the mode effective index for the Au-Si-Ti ($n_{im}$=0.368·i) structure is over twice as higher as the Au-Si-TiN structures ($n_{im}$=0.171·i and $n_{im}$=0.176·i); as a result, a twice as long Au-Si-TiN photodetector can provide the same absorption losses. At the same time, it can ensure much more hot carrier generation in a shorter distance from the metal-semiconductor interface.

When the gap between metals is further increased to g=500 nm, the absorption losses for a symmetric Au-Si-Au structure decrease as a result of lower electric field penetration into the Au (Fig. 7a). However, for the asymmetric structures (Fig. 7b and Fig. 8a, b), it can be observed that two separate modes propagate on each side of the metal-semiconductor interface. For the Au-Si-Ti structure, the real part of the mode effective index for the mode associated with the Au-Si interface is significantly higher than the Ti-Si mode effective index, while absorption is around 20 times smaller ($n_{eff}$=3.45-0.033·i for Au-Si and $n_{eff}$=3.27-0.656·i for Ti-Si). On the contrary, for the Au-Si-TiN structure the real part of mode effective index for both modes associated with the Au-Si and TiN-Si interfaces are very close to each other, with the absorption of mode associated with TiN-Si interface is 4-5 times higher compared to mode bounded to the Au-Si interface. Simultaneously, the electric field decays much faster at the TiN-Si interface compared to the Ti-Si interface, thus many more hot electrons can participate in a transition to the Si.

For TiN-Si interface, 90 % of the power is absorbed within 33 nm thick TiN area attached to Si. Compared to it, for Ti-Si interface this area increases to 50 nm [27]. Taking into account the electron mean free path in TiN and Ti being evaluated at 50 nm, the hot electron generated in TiN has higher probability to participate in transition to Si without interfacing a scattering compared to Ti. As a result, the internal quantum efficiency increases calculated at 35 % can be achieve.

## V. Dark Current

Apart from the responsivity and bandwidth, another important figure of merit of the photodetector is the signal-to-noise ratio (SNR) [28] defined as SNR = $i_s^2/i_n^2$ where $i_s$ and $i_n$ are the signal and noise currents, respectively. It is highly desired to enhance the signal while keeping the noise at low level. One solution to achieve a high SNR is by reducing the dimensions of the active Schottky junction area. Another solution is through optimizing the Schottky barrier between metal and semiconductor that should be as close as possible to the optimal value of ~ 0.697eV at telecom wavelength of 1550 nm (~ 0.8eV) that is calculated from $\Phi_{Bopt} = h\nu - 4kT/e$ [28]. Thus, the Schottky barrier height between TiN and p-doped Si being calculated at $\Phi_B$=0.67 eV based on our measurements is almost perfect when compared with the optimal value of $\Phi_{Bopt}$=0.697 eV for an ideal diode [28]. As a result, the p-Si-TiN provides a Schottky barrier height of 0.67 eV that it is much closer to the ideal value compared to p-Si-Ti or p-Si-Au contacts that were measured at 0.61 eV and 0.32 eV, respectively [28].

## VI. Conclusion

We propose titanium nitride as an alternative material for application in waveguide-integrated photodetectors. In addition to its CMOS-compatibility with standard fabrication technology, TiN offers superior electrical properties in terms of the Schottky barrier height, calculated at 0.67 eV for a junction formed between TiN and p-doped Si, and 0.35 eV for a junction between TiN and n-doped Si. The value of 0.67 eV is very close to the optimal Schottky barrier height of 0.69 eV for an operating wavelength of 1550 nm (~0.8 eV) that ensures very high signal-to-noise ratio performances. Simultaneously, titanium nitride offers superior optical properties for photodetection purposes – reasonably high negative real part of permittivity and reasonably high imaginary part enabling a high absorption and reduced penetration depth compared to Ti. All of those properties make titanium nitride a favorable material for fabrication of on-chip photodetectors.

## References

[1]. D. A. B. Miller, "Attojoule Optoelectronics for Low-Energy Information Processing and Communications," J. of Lightw. Technol. 35(3), 346-396, 2017.

[2]. L. C. Kimerling, D-L Kwong, and K. Wada, "Scaling computation with silicon photonics," MRS Bulletin 39, 687-695 (2014).

[3]. D. Thomson at el., "Roadmap on silicon photonics," J. of Optics 18, 073003, 2016.

[4]. S. Assefa, F. Xia. And Y. A. Vlasov, "Reinventing germanium avalanche photodetector for nanophotonic on-chip interconnects," Nature 464, 80-85, 2010.

[5]. S. Muehlbrandt, A. Melikyan, T. Harter, K. Kohnle, A. Muslija, P. Vincze, S. Wolf, P. Jakobs, Y. Fedoryshyn, W. Freude, J. Leuthold, C. Koos, and M. Kohl, "Silicon-plasmonic internal-photoemission detector for 40 Gbit/s data reception," Optica 3(7), 741-747, 2016.

[6]. I. Goykhman, U. Sassi, B. Desiatov, N. Mazurski, S. Milana, D. de Fazio, A. Eiden, J. Khurgin, U. Levy, and A. C. Ferrai, "On-Chip Integrated, Silicon-Graphene Plasmonic Schottky Photodetector for High Responsivity and Avalanche Photogain," Nano Lett. 16(5), 3005-3013, 2016.

[7]. I. Goykhman, B. Desiatov, J. Khurgin, J. Shappir, and U. Levy, "Locally Oxidized Silicon Surface-Plasmon Schottky Detector for Telecom Regime," Nano Lett. 11(6), 2219-2224, 2011.

[8]. I. Goykhman, B. Desiatov, J. Khurgin, J. Shappir, and U. Levy, "Waveguide based compact silicon Schottky photodetector with enhanced responsitity in the telecom spectral band," Opt. Express 20(27), 28594, 2012.

[9]. J. Gosciniak, Fatih B. Atar, Brian Corbett, and Mahmoud Rasras, "Plasmonic Schottky photodetector with metal stripe embedded into semiconductor and with a CMOS-compatible titanium nitride," Sci. Rep. 9, 6048 (2019).

[10]. N. Othman, and P. Berini, "Nanoscale Schottky contact surface plasmon "point detectors" for optical beam scanning applications," Appl. Optics 56(12), 3329-3334, 2017.

[11]. W. E. Spicer, "Photoemissive, Photoconductivite, and Optical Absorption Studies of Alkali-Antimony Compounds," Phys. Rev. 112, 114-122, 1958.

[12]. W. E. Spicer, "Negative affinity 3-5 photocathodes: Their physics and technology," Appl. Phys. 12, 115-130, 1977.

[13]. D. K. Gramotnev, and S. I. Bozhevolnyi, "Plasmonic beyond the diffraction limit," Nat. Photonics 4, 83-91, 2010.

[14]. W. L. Barnes, A. Dereux, and T. W. Ebbesen, "Surface plasmon subwavelength optics," Nature 424, 824-830, 2003.

[15]. "International Technology Roadmap for Semiconductors 2.0" 2015 Edition.

[16]. P. R. West, et al., "Searching for better plasmonic materials," Laser & Photonics Review 4(6), 795-808, 2010.

[17]. G. V. Naik, V. M. Shalaev, and A. Boltasseva, "Alternative plasmonic materials beyond gold and silver," Adv. Materials 25(24), 3264-3294, 2013.

[18]. J. Gosciniak, J. Justice, U. Khan. M. Modreanu, and B. Corbett. "Study of high order plasmonic modes on ceramic nanodisks," Opt. Express 25(6), 5244, 2017.

[19]. J. Gosciniak. J. Justice, U. Khan, and B. Corbett, "Study of TiN nanodisks with regard to application for Heat-Assisted Magnetic Recording," MRS Advances 1(5), 317-326, 2016.

[20]. W. Li, and J. G. Valentine, "Harvesting the loss: surface plasmon-based hot electron photodetection," Nanophotonics 6(1), 177-191, 2015.

[21]. T. P. Chen, T. C. Lee, C. C. ling, C. D. Beling, and S. Fung, "Current transport and its effect on the determination of the Schottky-barrier height in a typical system: Gold on silicon," Solid-State Electronics 36(7), 949-954, 1993.

[22]. T. Harter, et al., "Silicon-plasmonic integrated circuits for terahertz generation and coherent detection," Nat. Photonics 12, 625-633, 2018.

[23]. A. Naldoni, U. Guler, Z. Wang, M. Marelli, F. Malara, X. Meng, L. V. Besteiro, A. O. Govorov, A. V. Kildishev, A. Boltasseva, and V. M. Shalaev, "Broadband Hot-Electron Collection for Solar Water Splitting with Plasmonic Titanium Nitride," Adv. Optical Mater. 5, 1601031, 2017.

[24]. S. Ishii, S. L. Shinde, W. Jevasuwan, N. Fukata, and T. Nagao, "Hot Electron Excitation from Titanium Nitride Using Visible Light," ACS Photonics 3, 1552-1557, 2016.

[25]. Ch. Scales, I. Breukelaar, and P. Berini, "Surface-plasmon Schottky contact detector based on a symmetric metal stripe in silicon," Opt. Letters 35(4), 529-531, 2010.

[26]. A. Akbari, and P. Berini, "Schottky contact surface-plasmon detector integrated with an asymmetric metal stripe waveguide," Appl. Phys. Lett. 95, 021104, 2009.

[27]. S. Ishii, S. L. Shinde, W. Jevasuwan, N. Fukata, and T. Nagao, "Hot electron excitation from titanium nitride using visible light," ACS Photonics 3, 1552–1557, 2016.

[28]. M. Grajower, et al. "Optimization and experimental demonstration of plasmonic enhanced internal photoemission silicon schottky detectors in the mid-IR." ACS Photonics 4, 1015–1020, 2017.

[29]. S. Assefa, et al. "CMOS-integrated high-speed MSM germanium waveguide photodetector," Opt. Express 18, 4986–4999, 2010.